\documentclass[%
 reprint,
superscriptaddress,
 amsmath,amssymb,
 pra,
]{revtex4-1}

\usepackage{graphicx}
\usepackage{dcolumn}
\usepackage{bm}

\begin{document}


\title{Recoil-ion momentum spectroscopy of photoionization of cold rubidium atoms in a strong laser field}

\author{Renyuan Li}
\affiliation{Shanghai Advanced Research Institute, Chinese Academy of Sciences, 201210 Shanghai, China}
\affiliation{University of Chinese Academy of Sciences, 100049 Beijing, China}
\author{Junyang Yuan}
\affiliation{Shanghai Advanced Research Institute, Chinese Academy of Sciences, 201210 Shanghai, China}
\affiliation{University of Chinese Academy of Sciences, 100049 Beijing, China}
\affiliation{School of Physical Science and Technology, ShanghaiTech University, 201210 Shanghai, China}
\author{Xinya Hou}
\affiliation{College of Physics and Materials Science, Henan Normal University, 453007 Xinxiang, China}
\author{Shuai Zhang}
\affiliation{School of Physical Science and Technology, ShanghaiTech University, 201210 Shanghai, China}
\author{Zhiyuan Zhu}
\affiliation{Shanghai Advanced Research Institute, Chinese Academy of Sciences, 201210 Shanghai, China}
\affiliation{School of Physical Science and Technology, ShanghaiTech University, 201210 Shanghai, China}
\author{Yixuan Ma}
\affiliation{Shanghai Advanced Research Institute, Chinese Academy of Sciences, 201210 Shanghai, China}
\affiliation{School of Physical Science and Technology, ShanghaiTech University, 201210 Shanghai, China}
\author{Qi Gao}
\affiliation{Shanghai Advanced Research Institute, Chinese Academy of Sciences, 201210 Shanghai, China}
\author{Zhongyang Wang}
\affiliation{Shanghai Advanced Research Institute, Chinese Academy of Sciences, 201210 Shanghai, China}
\affiliation{School of Physical Science and Technology, ShanghaiTech University, 201210 Shanghai, China}
\author{T.-M. Yan}
\affiliation{Shanghai Advanced Research Institute, Chinese Academy of Sciences, 201210 Shanghai, China}
\author{Chaochao Qin}
\affiliation{College of Physics and Materials Science, Henan Normal University, 453007 Xinxiang, China}
\author{Yizhu Zhang}
\email{zhangyz@sari.ac.cn}
\affiliation{Shanghai Advanced Research Institute, Chinese Academy of Sciences, 201210 Shanghai, China}
\affiliation{Center for Terahertz Waves and College of Precision Instrument and Optoelectronics Engineering, Key Laboratory of Education, Tianjin University, 300072 Tianjin, China}
\author{Xincheng Wang}
\email{xinchengwang@gmail.com}
\affiliation{School of Physical Science and Technology, ShanghaiTech University, 201210 Shanghai, China}
\author{Matthias Weidem{\"u}ller}
\affiliation{Hefei National Laboratory for Physical Sciences at the Microscale and Shanghai Branch, University of Science and Technology of China, Shanghai 201315, China}
\affiliation{CAS Center for Excellence and Synergetic Innovation Center in Quantum Information and Quantum Physics, University of Science and Technology of China, Shanghai 201315, China}
\affiliation{Physikalisches Institut, Universit{\"a}t Heidelberg, Im Neuenheimer Feld 226, 69120 Heidelberg,Germany}
\author{Y.H. Jiang}
\email{jiangyh@sari.ac.cn}
\affiliation{Shanghai Advanced Research Institute, Chinese Academy of Sciences, 201210 Shanghai, China}
\affiliation{University of Chinese Academy of Sciences, 100049 Beijing, China}
\affiliation{School of Physical Science and Technology, ShanghaiTech University, 201210 Shanghai, China}
\affiliation{CAS Center for Excellence and Synergetic Innovation Center in Quantum Information and Quantum Physics, University of Science and Technology of China, Shanghai 201315, China}


\date{\today}

\begin{abstract}
We study photoionization of cold rubidium atoms in a strong infrared laser field using a magneto-optical trap (MOT) recoil ion momentum spectrometer. Three types of cold rubidium target are provided, operating in two-dimension (2D) MOT, 2D molasses, and 3D MOT with densities in the orders of $10^7$ atoms/cm$^3$, $10^8$ atoms/cm$^3$, and $10^9$ atoms/cm$^3$, respectively. The density profile and the temperature of 3D MOT are characterized using the absorption imaging and photoionization. The momentum distributions of Rb$^+$ created by absorption of two- or three-photon illuminate a dipole-like double-peak structure, in good agreement with the results in the strong field approximation. The yielding momentum resolution of $0.12 \pm 0.03$ a.u. is achieved in comparison with theoretical calculations, exhibiting the great prospects for the study of electron correlations in alkali metal atoms through interaction with strong laser pulses.
\end{abstract}

\pacs{}
\maketitle


\section{\label{sec:Introduction,level1}Introduction}
In the past few decades, the COLTRIMS (cold target recoil ion momentum spectroscopy) \cite{0034-4885-66-9-203,DORNER200095}, which enables measurement of recoil ion momentum in $4\pi$ solid angle with a high resolution for investigations of dynamical interaction between atoms/molecules and various projectiles, became one of the most fruitful experimental techniques in atomic, molecular and optical  physics. In the COLTRIMS, the targets are limited to the species of atoms/molecules in gas phase and volatilizable liquid, normally prepared by a supersonic jet. The usage of COLTRIMS-like apparatus is obstructive for alkali and alkaline-earth atoms due to their solid phase at room temperature. The heating of alkali atoms to gas phase gives rise to the undesired broadening of ionic momentum distributions due to severe thermal motion, leading to a few ten times lower resolution than commonly-used COLTRIMS.

Meanwhile, the hydrogen-like structure and low first ionization energy of alkali atoms open a new dimension to understand the strong-field ionization dynamics. For instance, time-resolved holography of photo-electrons reveals rich intra- and inter-cycle interferences \cite{Huismans61}, where metastable xenon atoms were prepared. Instead of noble gas, alkali atoms are ideal targets for the study of photoelectron interferences, particularly for intermediate-involved states in the far- and mid-infrared laser fields since alkali atoms can be easily prepared on various initial states with lasers. Besides, many amazing physical phenomena in the strong laser field, for instance, high harmonic generation (HHG) \cite{RevModPhys.81.163,KOHLER2012159}, above threshold ionization \cite{BECKER200235,PhysRevLett.110.013001}, tunneling ionization \cite{RN80,Eckle1525,PhysRevLett.118.143203}, and sequential and nonsequential double ionizations \cite{PhysRevLett.48.1814,PhysRevLett.99.263003,RevModPhys.84.1011,PhysRevLett.112.073002,PhysRevLett.113.103001,PhysRevLett.115.123001}, are still hot topics in the noble gas. However, these studies are still lacking due to highly technical challenges for momentum distribution detection of alkali recoil ion. On the other hand, alkali atoms with the extremely low first ionization potential provide new sights for understanding the mechanisms mentioned above. Furthermore, for HHG \cite{PhysRevLett.84.2822,PhysRevLett.94.113906,PhysRevLett.114.143902}, Kramers-Henneberger transformation \cite{Morales16906,WEI2017240}, and nonlinear ionization \cite{RN79}, alkali atom with respective to noble gas has been found to be of unusual features in the strong laser field.

The magneto-optical trap (MOT) technique was introduced to the COLTRIMS, called as magneto-optical trap recoil ion momentum spectroscopy (MOTRIMS) \cite{DEPAOLA2008139}, for investigations of interesting issues mentioned above with unprecedented recoil-ion momentum resolution. Recently, the rubidium (Rb) atom target was cooled down to a temperature of 200 $\mu$K \cite{doi:10.1063/1.4795475}, that is almost five orders lower than supersonic expansion target with similar atomic mass. The photoassociation of ultracold Rb${_2}$ was thus observed and optimized with the pulse-shaping technique of femtosecond laser pulses \cite{PhysRevLett.100.233003,doi:10.1063/1.4738643,doi:10.1063/1.4795475}. In addition, MOTRIMS had been employed for studies of photoionization dynamics of lithium (Li) atom in the free electron laser and infrared strong laser field \cite{schuricke_strong-field_2011,PhysRevLett.103.103008,PhysRevA.83.023413}, and of the few-body dynamics of Li atom in the ion collision \cite{PhysRevLett.109.113202}.

In this article, a MOTRIMS of Rb atoms for strong-field studies is introduced. Three types of targets, i.e. 3D (three-dimensional) MOT, 2D molasses\cite{PhysRevLett.55.48} and 2D MOT, are prepared to provide targets with various densities and different initial states, allowing to adapt to tunable intensities and different wavelengths of strong-field lasers. The 2D MOT target can be further cooled with six laser beams to form 2D molasses. The 2D molasses give higher density and the mixture of Rb(5s) and Rb(5p) as the initial states. The 2D molasses can be further trapped with a magnetic quadrupole field to form 3D MOT, which has the highest density. Similar to that of 2D molasses, the initial states of 3D MOT are also a mixture of Rb(5s) and Rb(5p).The setup of Rb MOTRIMS apparatus is depicted in the following sections, and the temperature, density and momentum resolution of three types are characterized.

\section{\label{sec:Setup,level1}Experimental setup}
\begin{figure}
\includegraphics[width = 8cm]{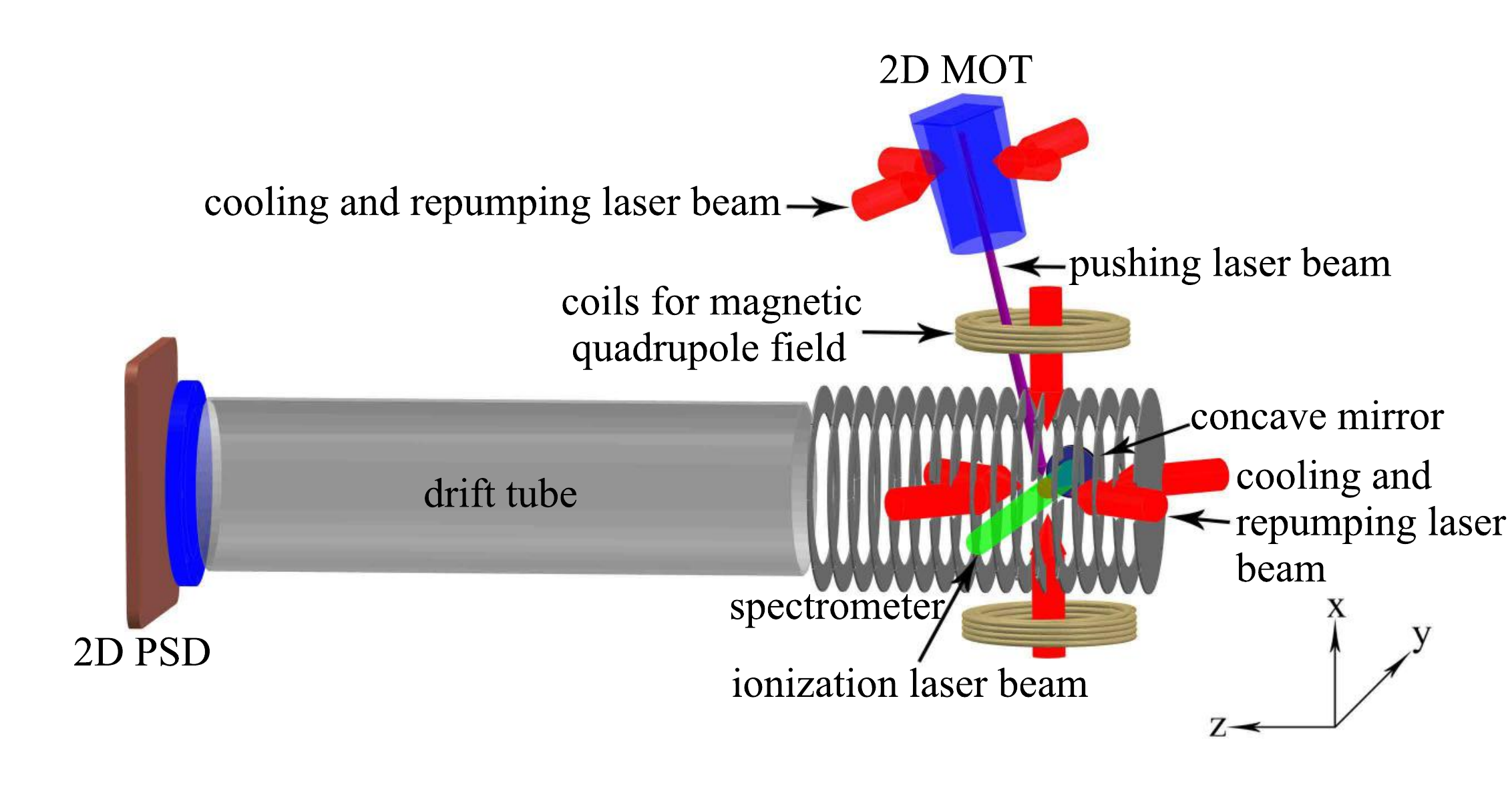}
\caption{\label{fig:setup} The sketch of the MOTRIMS. In the 2D MOT preparation part, the atoms can be pre-cooled in the $y$ and $z$ dimensions. In the target region part, the pre-cooled atoms will be further cooled in the $x$, $y$, $z$ dimensions and the momenta of recoil ion are detected with the RIMS part.}
\end{figure}

As indicated in Fig. \ref{fig:setup}, the setup mainly consists of four parts, dubbed as laser system, 2D MOT preparation region \cite{WEYERS199730,PhysRevA.80.013409,PhysRevA.66.023410,PhysRevLett.77.3331,PhysRevA.58.3891,PhysRevA.73.033415}, target region and recoil ion momentum spectrometer \cite{doi:10.1063/1.1775310,doi:10.1063/1.2994151}. Briefly, Rb atoms are pre-cooled in the 2D MOT preparation region. Then they are pushed into the main chamber, where they can be further cooled to form 2D molasses or 3D MOT target. The target is intersected with a femtosecond laser at the main chamber. The ions generated during the interaction are then extracted and detected by recoil ion momentum spectrometer.

The 2D MOT preparation region sits on top of the main chamber. A Rb dispenser is installed inside a glass cell, which can be heated to evaporate with a tunable current source. Before loading into the main chamber, the Rb atoms are pre-cooled in the glass cell by the cooling laser and the Rb density can be controlled by the heating current. Cooling laser beams with red detuning are split into four equal groups. The 2D MOT preparation part is separated from the main chamber via a differential tube, and the typical vacuum for the 2D MOT and main chamber are about $1 \times 10^{-8}$ mbar and $5 \times 10^{-10}$ mbar, respectively.

The target region is located at the main chamber, where the Rb atoms can be further cooled down by three pairs of orthogonal cooling lasers (2D molasses), with optional magnetic quadrupole field for trapping (3D MOT). A femtosecond laser of 800 nm was used to ionize the target, which offers a tool for characterization of targets and for further scientific investigations. Besides, the rubidium beam from 2D MOT preparation can also be used directly as a target for experiments of high ionization rate, or when the effect of the excited states have to be carefully considered

The recoil ions generated at the target region are extracted and detected by a standard RIMS (recoil ion momentum spectroscopy) method. MCP (micro channel plate) together with delay-line anode detector are used to record the arriving time and positions of each recoil ion, by which momentum vector and kinetic energy of recoil ions can be reconstructed.

\subsection{\label{sec:laser,level2}Laser system}

\begin{figure}
\includegraphics[width = 8cm]{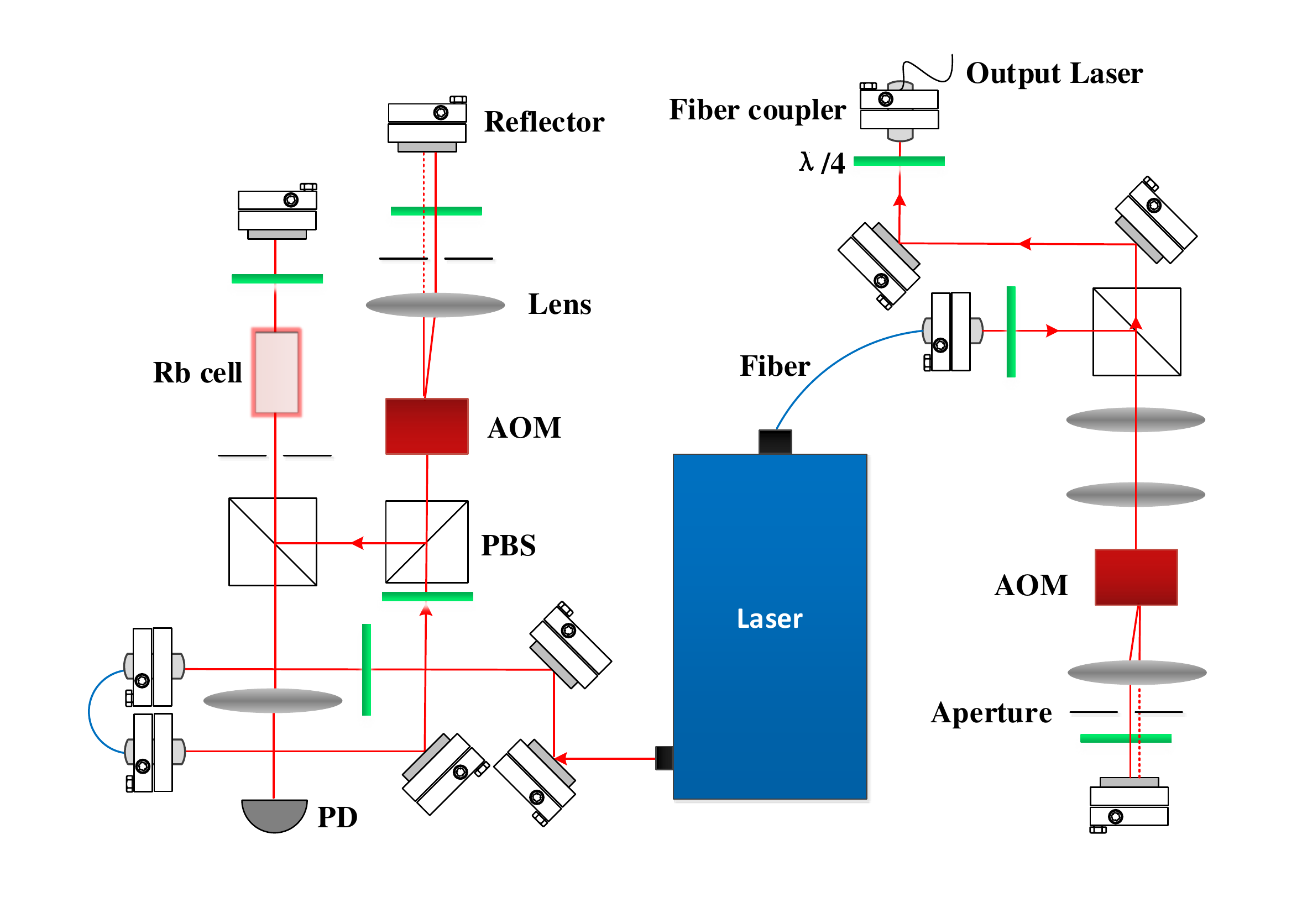}
\caption{\label{fig:laserSystem} Scheme of the laser system. AOM: acousto-optical modulators, PBS: polarization beam splitter, PD: Photo-Diode}
\end{figure}

A laser system is required to provide beams for cooling, pushing, imaging and repumping of Rb atoms. The lasers with the center wavelength of 780 nm for repumping and cooling, pushing, and imaging are locked on two different crossovers. Since the methodologies are quite similar, only one schematic design of the laser is presented in Fig. \ref{fig:laserSystem}.

Laser beam on the left side of Fig. \ref{fig:laserSystem} passing a Rb vapor cell is used for Doppler-free saturation spectroscopy \cite{doi:10.1119/1.18457} to stabilize the laser frequency. Acoustic-optical modulators (AOM) in a double-pass configuration shift the frequencies by $2 \times 80$ MHz for cooling, pushing and imaging beams, and by $2 \times 110$ MHz for repumping beam, where the laser beams are locked on the $F' = 3/4$ crossover (cooling, pushing and imaging) and $F' = 2/3$ crossover (repumping). With this kind of  optical design the working frequency of the laser beams on the right side can be stabilized.

The branch on the right side is a typical arrangement for cooling, pushing and imaging lasers with different detuning by AOM with the double-pass configuration. The imaging beam with a power of about 20 $\mu$W is set on the resonant transition of $|5\text{S}_{1/2}, F = 3, m_F = 3 \rangle \to |5\text{P}_{3/2}, F'=4, m_{F'} = 4 \rangle$. It is used in an absorption imaging experiment to measure the temperature and the density of 3D MOT target, described in detail in Section \ref{sec:Characterization,level1}, and whereas the pushing beam with a power of 0.2 mW is red-detuned by $-$2 $\times$ 12.5 MHz. A power of 190 mW in total (130 mW for the four groups of 2D MOT preparation and 60 mW for three pairs of beams of 2D molasses or 3D MOT) with red-detuning of $\delta$ = $-$2$\Gamma$ is needed. $\Gamma$ = 6 MHz is the natural linewidth of 5s5p, which determines the doppler temperature of about 150 $\mu$K.

Besides, the repumping beam with a power of 16 mW in total (12 mW for 2D MOT preparation and 4 mW for 3D moalsses or 3D MOT) is on resonant of the $|5\text{S}_{1/2}, F = 2, m_F = 2 \rangle \to |5\text{P}_{3/2}, F'=3, m_{F'} = 3\rangle$ transition. It can pump the Rb atoms back into the cooling cycle, avoiding accumulation of the dark state \cite{PhysRevA.53.1702,PhysRevLett.70.2253}. The output laser beams are then transported by polarization-maintained single mode optical fibers to the 2D MOT preparation and target region.

\subsection{\label{sec:2D,level2}2D MOT preparation}

The 2D MOT preparation part is used to provide a pre-cooled Rb beam for target region operation. To provide such an atom beam, normally, Zeeman slower and 2D MOT are adopted. However, comparing with 2D MOT, a Zeeman slower has two main disadvantages. One problem is that the Zeeman slower has cooling effect only in one direction, thus inferring a large spread in transverse velocities. On the other hand, for Zeeman slower, either the laser is on resonance or the magnetic field has to be non-zero at the end.
This means that at the target region, the 3D MOT for example, may be disturbed either by the resonant light or by the magnetic field of Zeeman slower \cite{doi:10.1063/1.1820524}. Here, we choose to use 2D MOT for preparation of a pre-cooled Rb beam leading to much more compact in the system geometry structure.

The Rb atoms generated in the glass cell are cooled with cooling lasers in two perpendicular directions associated with a quadrupole magnetic field, thus dubbed as 2D MOT. The cooling directions of $z$ and $y$ are perpendicular to the pushing beam direction as indicated in Fig. \ref{fig:setup}. To increase the target density in the target zone, four identical cooling regions along the pushing beam direction are designed with a total length of 80 mm. The cooling and repumping lasers mentioned above are split into four beams accordingly by PBS (polarization beam splitter) with a 1/e$^2$ diameter of 20 mm. On the other hand, the quadrupole magnetic field can be generated by a pair of anti-aligned permanent magnet. In our setup, six pairs of such magnets are aligned along the pushing beam, covering all the four cooling regions. Each piece of magnet can be adjusted independently to align the four regions. All the optical fiber couplers of the lasers and permanent magnets are fixed on a metal cage which is mounted around the glass cell.

After the 2D cooling in the glass cell, the Rb atoms will be pushed into the target region through a differential tube with diameter of 800 $\mu$m by a pushing beam with a 1/e$^2$ diameter of 1 mm set at one end of the glass cell. The pushed atom beam can interact directly with ionization laser beam as atom target, dubbed as 2D MOT target.

\subsection{\label{sec:3D,level2}Target region}

The target region is inside the main chamber, where three types of target can be used for various experiments. The space charge effect might occur due to large photoionization cross section of Rb atom, particularly for the creation of Rb$^+$ when the intensity of the femtosecond laser is high. This results in a deterioration of the momentum resolution. In this case, 2D MOT target with low density is expected to be a better option for Rb$^+$ detection. As an additional feature, because of the short lifetime of the excited state, 2D MOT provides a target with pure ground state, and thus these results from 2D MOT can be compared with these from 2D molasses or 3D MOT to study the effect of the excited state via detection of photoelectron in the future. Additional advantage is that the magnetic quadrupole field for trapping atoms in 3D MOT is not needed so that the detection efficiency increases significantly without taking time for switching on and off the magnetic field.

Alternatively, the Rb atoms can be further cooled with cooling and repumping lasers in three orthogonal dimensions to form 2D molasses target. The cooling and repumping lasers are combined in an optical fiber, split into three beams and conduced to the main chamber. Each beam comes from an optical coupler passes through a 1/4 $\lambda$ plate, a view port and then get into the target region to cool the Rb atoms. The remaining beam propagates though another viewport and again a 1/4 $\lambda$ plate and then get reflected by a plano mirror, thus realize cooling in the opposite direction. All laser beams intersect in the target region, where the 2D molasses target is formed.

It should be pointed out that the laser beam in the $x$ direction (see Fig. \ref{fig:setup}) can also be combined with an imaging laser for further absorption imaging experiments. Therefore, a PBS is set in front of the plano mirror, allowing cooling and imaging of the target at the same time.

Furthermore, for trapping cold atoms and increasing the target density, the magnetic quadrupole field generated from a pair of anti-Helmholtz coils shown in Fig. \ref{fig:setup} is adopted to form so-called 3D MOT. A pair of coils are mounted in a distance of 120 mm and each of them has the mean diameter is about 70 mm. The cooling water passing through the coils for maintaining the temperature in constant. 2D MOT, 2D molasses and 3D MOT are proper for different types of experiments and their profiles will be discussed in details later.

\subsection{Recoil ion momentum spectrometer}
At the target region, a femtosecond laser of 800 nm interacts with the Rb targets and creates recoil ions in various charged states with certain momentum. The recoil ion momentum spectrometer is a powerful instrument to detect the momentum of recoil ions with 4$\pi$ solid angle. The spectrometer in our setup consists of electrodes for generation of a uniform extraction field, a field-free drift tube for improvement of the resolution, and a MCP (micro channel plate) plus delay-line anode for momentum detection of individual recoil ion.

There are 34 pieces of ring electrodes made of stainless steel. Every electrode has a same thickness of 1 mm and the distances between adjacent electrodes are always 4 mm. The outer diameter of the electrodes is 100 mm and the inner diameter is 75 mm. At the far end from the detector, there is a repeller, i.e. a disk with a diameter of 100 mm and a thickness of 1 mm. The electrode closest to the detector is connected with a metal tube to provide a field-free drift region. The tube has a length of 670 mm and a inner diameter of 81 mm. By a bias voltage between the drift tube and the repeller, a uniform electric field is generated as all adjacent electrodes are connected by resistors of 1 M$\Omega$. For the photoionization experiment, if not specified, an electric field of 0.5 V/cm is set to balance between a better resolution of recoil ion momentum and a preferable efficiency.

A commercial Z-stack MCP detector with delay-line anode is used for the detection of recoil ions. The active area of the detector is about 80 mm diameter, and a position resolution of 0.1 mm can be realized. The time resolution of the recoil ion momentum spectrometer is about 1 ns, and a time range of 400 $\mu$s for TOF measurement is achieved.

\section{\label{sec:Characterization,level1}Characterization of the setup}

\subsection{Characterization of cold atom targets}

\begin{figure}
\includegraphics[width = 8cm]{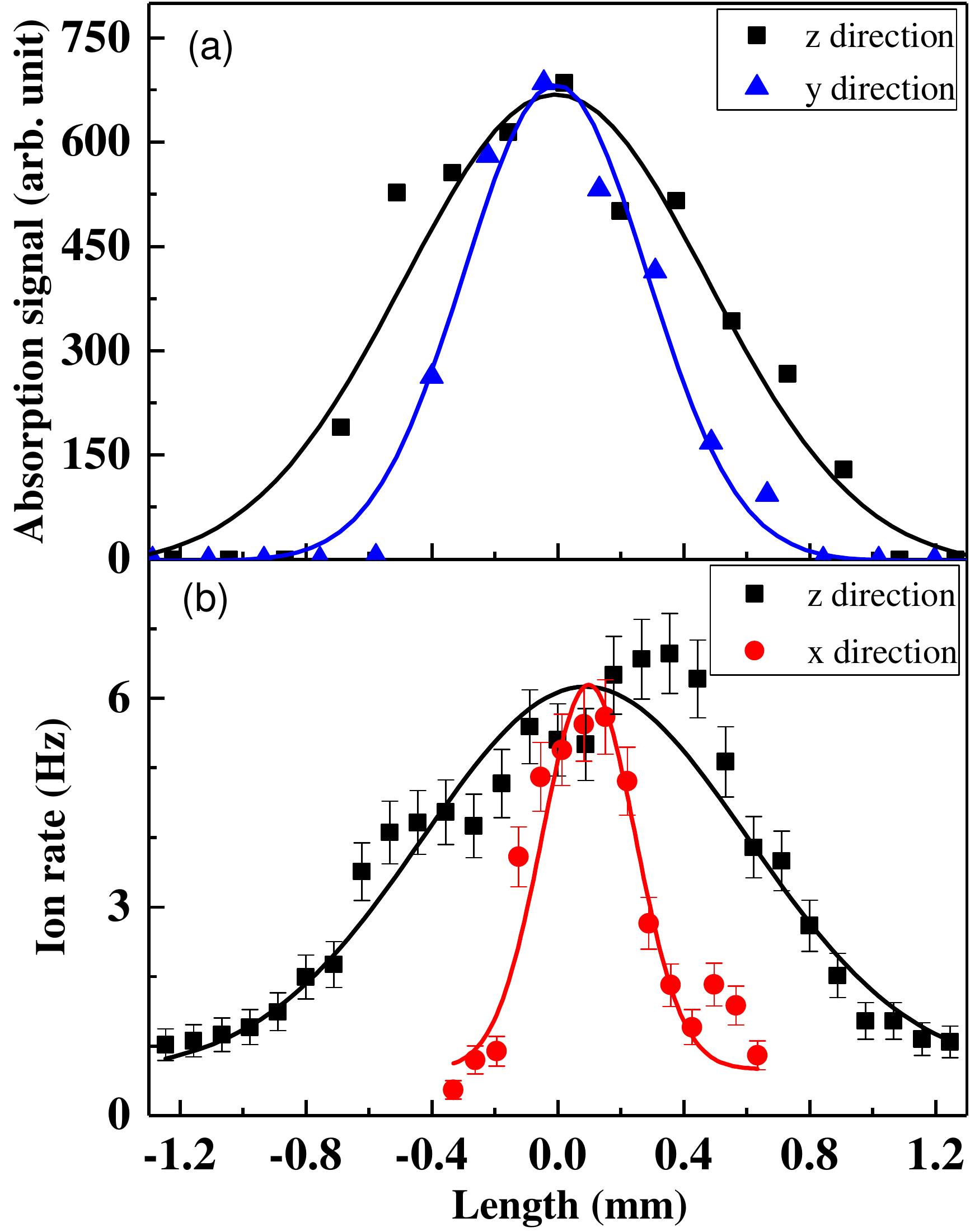}
\caption{\label{fig:motSize} Density profile of 3D MOT as a function of its central positions. Density distributions are given by absorption experiment in the $z$ (black square) and the $x$ (blue triangle) directions (a) and  by laser ionization in the $z$ (black square) and $y$ (red spot) directions (b). All curves are corresponding data fittings to Gaussian function.}
\end{figure}

The density profile of 3D MOT is measured as a function of its central position by absorption imaging method and photoionization approach given in Fig. \ref{fig:motSize}. Here, the imaging laser beam is transported from the top of the main chamber in the $x$ direction. A CMOS camera is set after the PBS for imaging 3D MOT in the $y$ and $z$ directions. From the FWHMs (full width at half maxima) of the Gaussian fits, the density distributions of 3D MOT are determined to be about 1.1 mm and 0.7 mm in the $y$ and $z$ directions, respectively. Taking into account imaging of pure background (no imaging laser beams and MOT), imaging with cooling lasers (no MOT beam) \cite{Ketterle99making}, and imaging with lasers and MOT, the number of cold atom trapped is estimated to be around $1.5 \times 10^6$ atoms. By switching off the cooling and repumping lasers together with the magnetic field and taking the absorption images at various times, the expansion of 3D MOT was measured to be 0.11 m/s corresponding to temperature of 130 $\mu$K in the $y$ and $z$ directions.

On the other hand, the relative density profiles of 3D MOT can also be obtained by photoionization via scanning the focus spot of the femtosecond laser in the position assuming that the ionization rate is proportional to the target density. Here, the wavelength of 800 nm, the repetition rate of 1 kHz, and the maximum power output of about 1 mJ per pulse were used. The laser intensities are adjusted using a calcite wollaston prism and a half-wave plate before entering the main chamber. Unfocused laser beam first goes through science chamber without ionizing the targets because of its extremely low intensity and then is focused back on the targets by a concave mirror with a focal length of 75 mm mounted on the opposite end of the incident femtosecond laser inside the main chamber. The focus spot in a 1/e$^2$ diameter of 20 $\mu$m and the Rayleigh length of 716 $\mu$m in the $y$ direction are achieved.  Rayleigh length may result in relatively worse spatial resolution, which can be realized in the momentum distributions below.

In present setup, the focusing lens is mounted on a multi-axis manipulator with a travel range of 75 mm in the femtosecond laser propagation direction ($y$ direction in Fig. \ref{fig:setup}) and 16 mm in the $x$ and $z$ directions. Unfortunately, the movement of manipulator cannot be controlled in the very precise step. Instead, we use magnetic field from two pairs of coils to move the 3D MOT in the $x$ and $z$ directions. Changing the current of the coils, the 3D MOT target moves almost linearly with scale coefficients of 1.04 mm/A in the $x$ direction and 0.63 mm/A in the $z$ direction. With a step of about 0.08 mm, the 3D MOT density profiles plotted in Fig. \ref{fig:motSize}(b) are scanned by counting ionization rate of Rb$^+$ as a function of position. On the basis of the Gaussian fits, the density distributions of the 3D MOT in the $z$ and $x$ directions are given by 1.22 mm and 0.35 mm in FWHM, respectively. With the number of cold atom and the size of 3D MOT mentioned above, the density is estimated to be about $5 \times 10^9$ atoms/cm$^3$. Furthermore, assuming linear dependence between Rb$^+$ count rate and the atomic density, the atomic densities of 2D MOT and 2D molasses are estimated to be $10^7$ atoms/cm$^3$ and $10^8$ atoms/cm$^3$, respectively.

\subsection{Recoil ions momentum distribution of cold atom targets}
One of the most important applications for Rb MOTRIMS is to study the electron correlation dynamics in the strong laser field by investigations of the momentum distribution for the recoil ions, which represents momentum sum of all outgoing electrons. In the strong laser field, multiphoton ionization and tunnel ionization can occur, where Keldysh parameter \cite{KELDYSH1965Ionization}, $\gamma = \sqrt{I_\text{p}/2U_\text{p}}$, is used to categorize both processes, i.e. $\gamma \gg 1$ for multi-photon ionization and $\gamma \ll 1$  for tunnel ionization. $I_\text{p}$ and $U_\text{p}$ indicate the atomic ionization potential and the ponderomotive potential, respectively.

\begin{figure}
\includegraphics[width = 8cm]{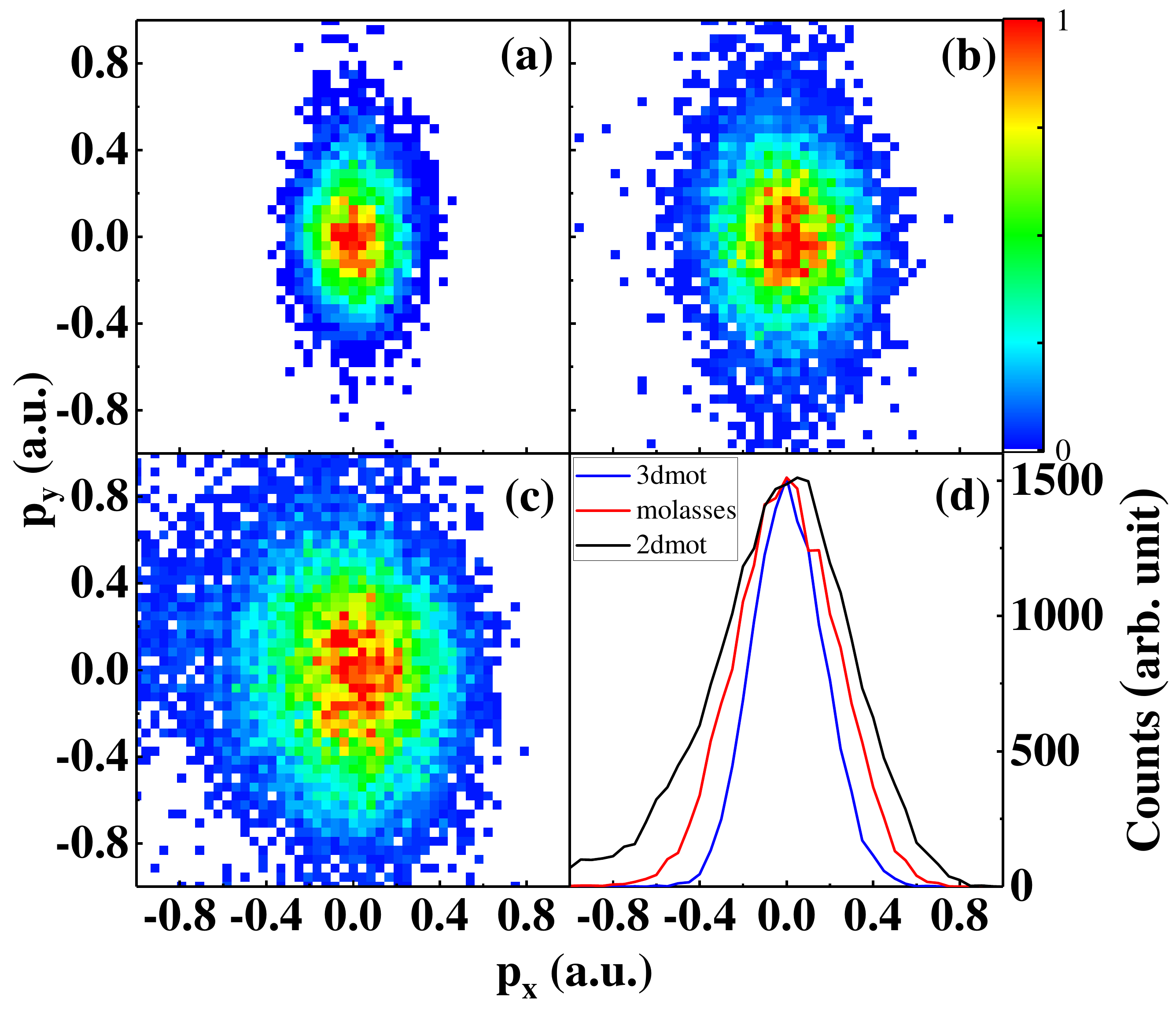}%
\caption{\label{fig:pxpy} Density plot of the measured Rb$^+$ momentum distributions for 3D MOT (a), 2D molasses (b), 3D MOT (c), and momentum distributions projected into the $x$ direction for three types of targets (d). The $x$- and $y$-axis indicate the direction perpendicular to the polarization plane and the propagation direction of the femtosecond laser beam, respectively}
\end{figure}

Density plot of the measured recoil-ion momentum distributions for Rb$^+$ in the $xy$-plane are presented in Fig. \ref{fig:pxpy}. Here, linear polarization of the ionizing femtosecond laser with 800 nm wavelength is set in the $z$ direction. The femtosecond laser intensities are estimated to be $2.0 \times 10^{10}$ W/cm$^2$, 1.0 $\times$ 10$^{11}$ W/cm$^2$ and $9.0 \times 10^{11}$ W/cm$^2$ for 3D MOT, 2D molasses and 2D MOT targets corresponding to Keldysh parameters of about 45, 19 and 6, respectively. This means that multiphoton ionization for creation of Rb$^+$ is dominating in all three cases. Note that, the low intensities are selected so that the space charge effect is negligible.

In the Fig. \ref{fig:pxpy}, one clearly find that momentum distributions of the measured Rb$^+$ recoil-ion in the $p_x=$ 0.40 a.u. (FWHM) and  the $p_y=$ 0.60 a.u. for 3D MOT are narrower than $p_x = 0.56$ a.u. and the $p_y = 0.70$ a.u. for 2D molasses and $p_x = 0.69$ a.u. and the $p_y = 0.99$ a.u. for 2D MOT. For the $x$ directions in 2D molasses and 2D MOT, the momentum distributions are essentially determined by the target temperature, whereas these in the $y$ direction, worse than the $x$ direction, are affected by the focusing Rayleigh length of the femtosecond laser. For a direct comparison, the normalized $p_x$ momentum components are plotted in Fig. \ref{fig:pxpy}(d).

\begin{figure}
\includegraphics[width = 7.7cm]{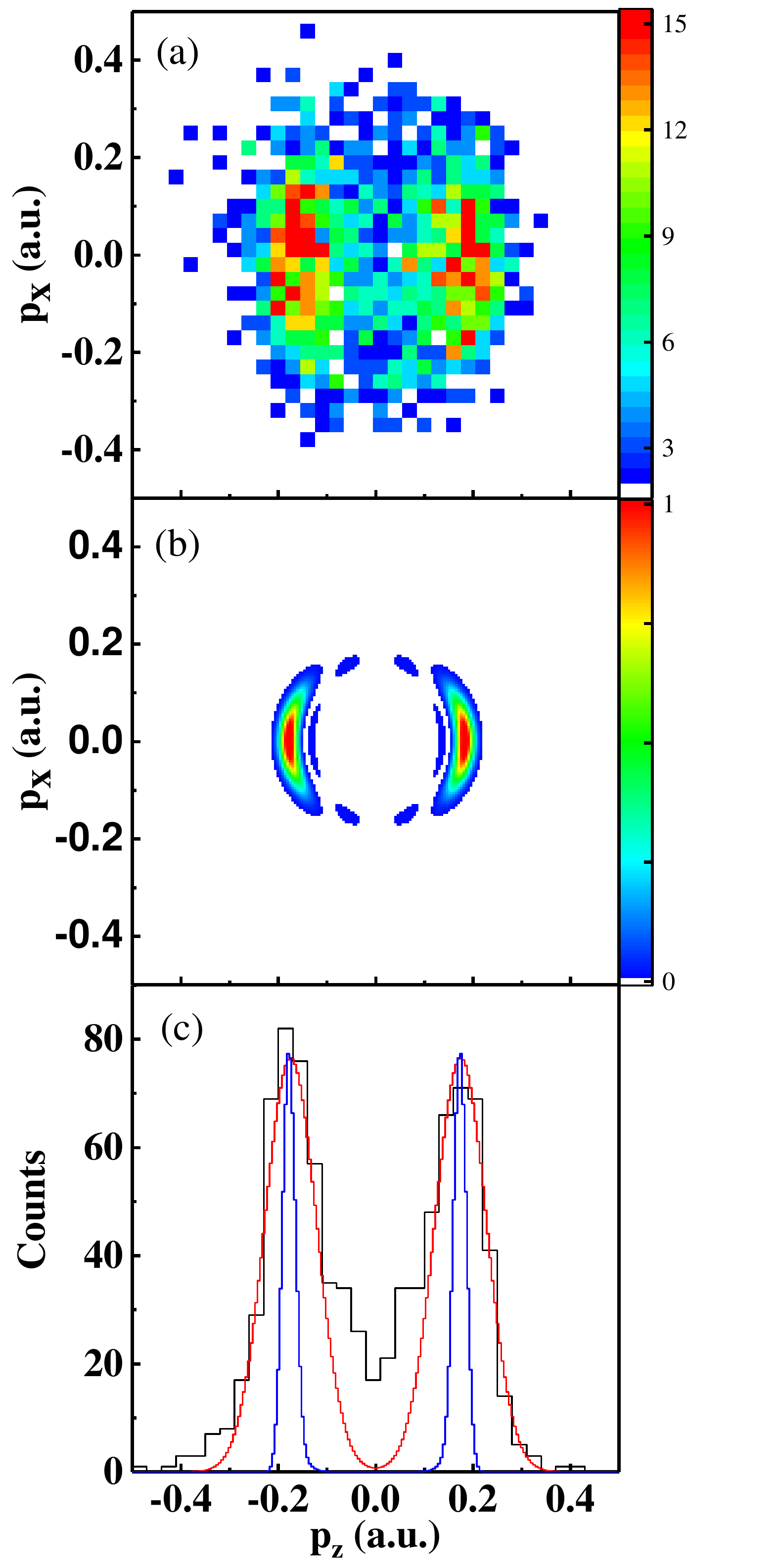}%
\caption{\label{fig:pxpz} Density plot of Rb$^+$ momentum distribution ($|P_y|<$ 0.1 a.u.) in the $z$ and $x$ directions for the measurement (a) and SFA calculation (b). The experimental data (black curve) and the calculation (blue curve) sliced with $\sqrt{p_x^2 +p_y^2} < 0.1$ a.u. are plotted in (c), where theoretical result (red curve) convoluted with $0.12 \pm 0.03$ a.u. is displayed in red line. }
\end{figure}

Density plot of Rb$^+$ momentum distribution together with the calculations in the $z$ and $x$ directions for 3D MOT target are displayed in Fig. \ref{fig:pxpz}, where $z$ represents the polarization direction. Present calculations are under the single-active electron approximation. The momentum distribution is evaluated by $w(\boldsymbol{p}) = |p| |M_{\boldsymbol{p}}|^2$ using the strong field approximation (SFA), where the ionization amplitude $M_{\boldsymbol{p}} = \int \mathrm{d}t \langle \boldsymbol{p} + \boldsymbol{A}(t') | \boldsymbol{r}\cdot \boldsymbol{E}(t') | \Psi_0 \rangle \exp[\mathrm{i} S(t')]$ and $S(t') = \int^{t'} \mathrm{d} t''[(\boldsymbol{p} + \boldsymbol{A}(t''))^2 / 2 + I_{\text{p}}]$ with $\boldsymbol{E}(t')$ and $\boldsymbol{A}(t')$ the electric field and vector potential of the laser field, respectively. The vector potential $\boldsymbol{A}(t')$ takes the $\sin^2$-envelope of 20 cycles (duration $\sim$ 26 fs), and the intensity of the laser field is $10^{10}$ W/cm$^2$.

In Fig. \ref{fig:pxpz}(a), experimental momentum distribution shows a dipole-like double-peak structure reproduced by our calculation shown in Fig. \ref{fig:pxpz}(b) and a nice agreement between experiments and calculations is found. The 3D targets with cooling laser constitute atoms mixed in the ground state and in the first excited state and therefore the recoil ions of Rb$^+$ are created by absorption of three photons and two photons, respectively:
\begin{eqnarray*}
\text{Rb}(^2\text{S}_{1/2}) + 3{h}\nu & = & \text{Rb}^+(^1\text{S}_0) + \text{e}^- \\
\text{Rb}(^2\text{P}_{3/2}) + 2{h}\nu & = & \text{Rb}^+(^1\text{S}_0) + \text{e}^- .
\end{eqnarray*}
According to the ionization energies of 4.18 eV and 2.58 eV from the ground state and the first excited state, respectively, both release an almost same excess-energy of $E_e = 0.47$ eV for the outgoing electron. This results in a recoil momentum of 0.19 a.u., which is good agreement with the observed and calculated results of $\pm 0.18$ a.u. displayed in Fig. \ref{fig:pxpz}. Here, two ionization progresses from the ground state and the first excited state cannot be distinguished only by kinetic energy of recoil-ion since the first excited energy of Rb pumped by cooling laser is almost same with single-photon energy of femtosecond laser. In future work, contributions of the ground state and the first excited state in 3D MOT target will be investigated by Rb$^+$ intensity dependence for the angular distribution and the ionization rate. Limited to present experimental momentum resolution, the substructure from the inner ring indicated by theoretical results cannot be distinguished clearly.

Taking a region of $\sqrt{p_x^2 +p_y^2} < 0.1$ a.u., the experimental momentum distribution of $p_z$ as well as theoretical result indicated with blue curve were plotted in Fig. \ref{fig:pxpz} (c). Again, a double-peak structure is illuminated more clearly and the positions of two peaks are in good agreement between experiment and theory.
For a direct comparison, theoretical result in red line is convoluted with a Gaussian function yielding an experimental momentum resolution of 0.12 a.u.. We also note that 2D molasses and 2D MOT targets show the same momentum resolution with 3D MOT in the $z$ (TOF) direction. Here, the momentum distributions in the $z$ direction is reconstructed with TOF of each event by considering the geometry and voltage settings of the spectrometer. Present momentum resolution is mainly limited to the uncertainties of extraction electric field and other stray fields. By changing the polarization of femtosecond laser to the $x$ direction, i.e., along the 2D beam movement direction shown in Fig. 1,  a similar double-peak structure is also observed in three targets. By this way, the momentum resolutions extracted along the $x$ direction are less than 0.2 a.u. for 3D MOT but more worse for 2D molasses and 2D MOT where the velocity distributions make the temperatures up to a few mK high.

It should also be pointed out that it is more convenient to use 2D molasses due to its accepted resolution mentioned above since  quadrupole magnetic field for trapping atoms is not applied so that electrons can be detected with the high detection efficiency. Imaging electron momentum distributions is being proposed. We also note that very recently a MOTRIMS-like design for lithium was introduced in Ref. \cite{PhysRevA.97.043427}.

\section{\label{sec:Conclusion,level1}Conclusions}

New magneto-optical trap (MOT) recoil ion momentum spectroscopy for studies of Rb atom in the strong infrared laser field was developed, in which three types of cold Rb target are provided, i.e., 2D MOT, 2D molasses, and 3D MOT. In the present design, three targets can provide  densities in the orders of $10^7$ atoms/cm$^3$, $10^8$ atoms/cm$^3$, and $10^9$ atoms/cm$^3$, respectively. The lowest temperature for 3D MOT target up to 130 $\mu$K is achieved. The cold targets with various densities are accessible for different investigations considering the effects of space charge and excited states involved. The momentum distributions of Rb$^+$ created by the multi-photon absorption for three types of atom targets are measured and a dipole-like double-peak feature is well reproduced with the strong field approximation. With the help of calculations in the strong field approximation, the momentum resolution of $0.12 \pm 0.03$ a.u. along the time-of-flight is achieved. Present first observation for momentum distributions demonstrates that MOTRIMS apparatus combined with the femtosecond laser pulses provides a powerful and new tool for investigations of the strong-ionization dynamics in alkali atoms.

In the near future, electron detection will be developed, allowing for ion and electron detections in coincidence. Considering the multi-functions for imaging ions, electrons, and photons, present MOTRIMS for alkali atoms are of great potentials for the study of Rydberg multi-body interactions as well as the strong-correlations in the cold plasma.

\section{Acknowledgment}

We acknowledge financial supports by the Instrument Developing Project of the Chinese Academy of Sciences (YZ201537), National Natural Science Foundation of China (NSFC) (91636105, 11420101003, 61675213, 11604347), and Shanghai Sailing Program (16YF1412600). We are grateful for numerous discussions with Alexander Dorn, Bastian H\"oltkemeier, Xueguang Ren, Jiefei Chen and Peng Chen.

\bibliographystyle{apsrev4-1}
\bibliography{main}

\end{document}